\documentstyle[12pt]{article}

\if@twoside\oddsidemargin15mm
\else\oddsidemargin15mm\evensidemargin15mm\marginparwidth15mm\fi%
\newsavebox{\uuunit}
\sbox{\uuunit}
    {\setlength{\unitlength}{0.825em}
     \begin{picture}(0.6,0.7)
        \thinlines
        \put(0,0){\line(1,0){0.5}}
        \put(0.15,0){\line(0,1){0.7}}
        \put(0.35,0){\line(0,1){0.8}}
       \multiput(0.3,0.8)(-0.04,-0.02){12}{\rule{0.5pt}{0.5pt}}
     \end {picture}}

\newcommand{\dr}{\raise.3ex\hbox{$\stackrel{\leftarrow}{\delta }$}}
\newcommand{\dl}{\raise.3ex\hbox{$\stackrel{\rightarrow}{\delta}$}}
\newcommand{\beq}{\begin{equation}}
\newcommand{\eeq}{\end{equation}}
\newcommand{\ddr}{\raise.3ex\hbox{$\stackrel{\leftarrow}{d}$}}
\newcommand{\ddl}{\raise.3ex\hbox{$\stackrel{\rightarrow}{d}$}}

\def\gtwid{\raise.3ex\hbox{$>$\kern-.75em\lower1ex\hbox{$\sim$}}}
\def\ltwid{\raise.3ex\hbox{$<$\kern-.75em\lower1ex\hbox{$\sim$}}}

\newcommand{\nit}{\noindent}
\newcommand{\vs}{\vspace{2ex}}

\begin{document}

\pagestyle{empty}

\begin{flushright}
NIKHEF/95-032 \\
DAMTP-95/36 \\
hep-th/9507046
\end{flushright}

\begin{center}
\Large{{\bf NEW SUPERSYMMETRY OF THE MONOPOLE}} \\
\vspace{3ex}

\large{ F.\ De Jonghe, A.J.\ Macfarlane$^*$, K.\ Peeters and J.W.\ van Holten}
\\
\vspace{3ex}

\large{ NIKHEF, Postbus 41882, 1009 DB Amsterdam,  The Netherlands} \\
\vspace{5ex}

June 1995 \\
\vspace{7ex}

\small{{\bf Abstract}} \\
\end{center}
\vs

\nit
\small{
The non-relativistic dynamics of a spin-1/2 particle in a monopole field
possesses a rich
supersymmetry structure. One supersymmetry, uncovered by d'Hoker and Vinet,
is of the standard type: it squares to the Hamiltonian. In this paper we
show the presence of another supersymmetry which squares to the
Casimir invariant of the full rotation group. The geometrical origin of
this supersymmetry is traced, and its relationship  with the constrained
dynamics of a spinning particle on a sphere centered at the monopole is
described. }
\vspace{50ex}

\nit
\small{
$^*$ On leave of absence from DAMTP, University of Cambridge, Silver Street,
Cambridge CB3 9EW, UK, and St. John's College, Cambridge. }

\newpage

\pagestyle{plain}
\pagenumbering{arabic}

\nit
{\bf 1.} This paper discusses spin-$\frac{1}{2}$ particles in a magnetic
monopole background. The model is described by the Hamiltonian
 \begin{equation}
      H = \frac{1}{2} ( \vec{p} - e \vec{A} )^2 -e \vec{B} . \vec{S}\,   ,
  \end{equation}
\nit
where all vectors are 3-dimensional. In particular $\vec{S} = \vec{\sigma}/2$,
$ \vec{B} = g \vec r/r^3$, and the vector potential $\vec{A}$ is to be defined
patchwise in the well-known way. It is more than ten years now since it was
observed that this model posesses a hidden supersymmetry \cite{dhoker}.
This supersymmetry is of the so-called $N = \frac{1}{2} $ type \cite{Saki} -
\cite{AJM}  in which there is a single hermitian supercharge $Q$ such that
$Q^2= H$. It is unusual in that the phase space of its fermionic sector is of
odd dimension: this involves three hermitian Majorana fermions $ \psi_i$,
$i=1,2,3$. These satisfy $\{ \psi_i , \psi_j \} = \delta_{ij}$ and admit the
natural representation $ \psi_i =  \sigma_i/\sqrt{2}$, in terms of
Pauli matrices, which gives rise to the description \cite{martin} of the spin
of the monopole.  In the rest of this paper, we will work at the level of the
quantum theory, i.e.\ after the transition from Poisson brackets to canonical
(anti)commutators. However, we will not use the specific representation of the
Grassmann coordinates in terms of Pauli matrices that was mentioned above,
since this can
obscure the difference between Grassmann even  and Grassmann odd operators.

In this paper, we show that the dynamics of a spin-$ \frac{1}{2}$ particle in a
magnetic monopole field admits a larger hidden supersymmetry structure. Indeed,
the theory possesses an additional supercharge $ \tilde Q$ which obeys
 \begin{equation}
    \{ Q , \tilde Q\} = 0 \, ,
    \label{antiQ}
 \end{equation}
\nit
so that $[\tilde Q , H] = 0 $ is evident. Also we have
  \begin{eqnarray}
      \label{squareK}
     \tilde Q^2 & = & K   \nonumber     \\
     K  & = & \frac{1}{2} [ \vec{J} {}^2 - e^2 g^2 + \frac{1}{4} ] \, ,
   \end{eqnarray}
\nit
where $\vec{J}$ is the total angular momentum of the monopole system,
\begin{equation}
   \vec{J} = \vec{r} \times \left( \vec{p} - e \vec{A} \right) + \vec{S}
   - eg \frac{\vec{r}}{r} \, .
\end{equation}
\nit
The operator $\tilde Q$ ---which coincides up to a constant shift with the
operator $A$ in \cite{dhoker}--- plays an important role in the
determination of the spectrum of the model. To see the equivalence between
$\tilde Q$ and $A$, one has to use the representation of the Grassmann
coordinates
by Pauli matrices, and, in  this way, one loses the insight that $A$ is
in fact a Grassmann odd operator. In the original analysis, $A$  was introduced
only
on algebraic grounds, and its geometrical interpretation remained unclear.
Here we explain that it corresponds to a new type of supersymmetry of
the problem, and our way of obtaining this extra supercharge shows the geometry
of the monopole configuration to be considerably richer than anticipated. In
particular, since the new supercharge $\tilde Q$ squares to essentially
$\vec{J}{}^2$,
our analysis shows that the full supersymmetry algebra of the monopole is in
fact a {\it non-linear algebra}.
\vs

\nit
{\bf 2.} Our discussion is based, firstly, on a general  method \cite{GRH}
which uses Killing-Yano tensors \cite{KY} to generate additional
supersymmetries in a theory which already possesses some known supersymmetries
of a standard type, and, secondly, on the realisation that vital theoretical
information can stem from the employment, where appropriate, of fermionic phase
spaces of odd dimension.  We have seen that the last circumstance does apply to
the monopole. Also, it is true that the flat background of the monopole
admits the simplest available non-trivial example of a Killing-Yano tensor,
viz.
\begin{equation}
f_{ij} = \varepsilon_{ijk}x_k   \, .
\label{ourKY}
\end{equation}
\nit
This satisfies trivially the conditions generally obeyed \cite{CM,GRH} by such
a tensor, namely,
\begin{eqnarray}
     f_{ij}  & =& - f_{ji}  \, ,  \nonumber   \\
      f_{ij,k} +  f_{ik,j}& = &  0    \, .
\end{eqnarray}
\nit
Further, we can use it, as in \cite{GRH}, to define a hermitian  supercharge
$\tilde Q = \tilde Q^\dagger$, which obeys (\ref{antiQ})  and (\ref{squareK}).
We develop these matters below starting from the superfield formulation. Our
treatment may be compared with the recent work of \cite{tani}, which added
electromagnetic interactions to the curved space analysis of spinning
particles in $d$ dimensions in \cite{GRH}. Although \cite{tani} does not
address the special properties and simplifications that occur in the specific
case of $d=3$, our work evidently has links with the general results found
there.

We use scalar superfields
\begin{equation}
   \Phi_i = x_i + i \theta\psi_i,  \, \hspace{3em} i=(1,2,3),
\end{equation}
\nit
involving one real Grassmann variable such that $\theta^2 = 0$, $\theta
= \theta^*$, a real coordinate $x_i$ and Majorana fermions $\psi_i$. The
most general Lagrangian we can build is of the form
\begin{equation}
 S = \int dt  \,\, d\theta  \left[ {i\over 2} \dot{\Phi_i} D \Phi_i  +
 ie D\Phi_i A_i(\Phi) +
 {i \over 6} k \,  \varepsilon_{ijk}  D\Phi_i D\Phi_j D\Phi_k \right]  \,   ,
\label{genac}
\end{equation}
\nit
where the operator $D = \partial_\theta - i\theta \partial_t$.
This contains kinetic, electromagnetic and torsion terms, but no pure
potential term can be built without the use of  spinor superfields (see e.g.
\cite{AJM} for more details on this point).  However, we do not need the
latter here and will simplify (\ref{genac}) by taking
the coupling constant for the
torsion term to be zero at first.

Writing the action out in component fields one obtains
\begin{equation}
  L = {1 \over 2} \dot{x}_i \dot{x}_i + {1 \over 2}  i \psi_i \dot{\psi}_i
  + e \dot{x}_i A_i(x) - {1 \over 2}  i e F_{ij} \psi_i\psi_j \, ,
\end{equation}
\nit
where now $A_i(\vec{r} )$ is seen to be the vector potential of the magnetic
field $\vec{B} $, with components $B_i =  \varepsilon_{ijk} A_{k,j} =
\frac{1}{2}\, \varepsilon_{ijk} F_{jk}$.
As usual \cite{dhoker},
\begin{eqnarray}
\label{basbra}
         p_i & = & \dot{x}_i + eA_i  \, , \nonumber \\
         \left[ x_i,p_j \right] & = & i \delta_{ij} \, , \nonumber \\
         \{\psi_i,\psi_j\} &= & \delta_{ij} \, ,
\end{eqnarray}
\nit
and $H = Q^2$ for the supercharge
\begin{equation}
         Q = (p_i - e A_i(x) ) \psi_i       \, .
\label{supe1}
\end{equation}
\nit
Following the procedure of \cite{GRH}, we seek a new supercharge $\tilde Q$,
which obeys (\ref{antiQ}) and hence describes a new supersymmetry of $H$, in
the form
\begin{equation}
\label{ourQ}
   \tilde Q =  (p_i - e A_i)  f_{ij} \psi_j   +i {b \over 6}
\varepsilon_{ijk}\psi_i\psi_j\psi_k   \, ,
\end{equation}
\nit
with the tensor $f_{ij}$ given by (\ref{ourKY}), and some parameter $b$ to be
determined by demanding that (\ref{antiQ}) holds. Using the basic commutators
(\ref{basbra}), one finds that for arbitrary $\vec{A}$,
\begin{equation}
    \{ Q , \tilde Q\} = -2 (1 + b) (\vec{p} - e \vec{A}) \cdot \vec{S} + e
\vec{B} \times \vec{r}  \cdot  \vec{S} \, .
\end{equation}
\nit
where $S_i = - {i \over 2}\varepsilon_{ijk} \psi_j\psi_k$ defines the spin
operator \cite{dhoker}. We chose $b=-1$, and see that the anticommutator
vanishes for radial $\vec{B}$, and therefore for
the monopole, for which
\begin{equation}
   \vec{B} = g \frac{\vec r}{r^3 }   \, .
\end{equation}
The condition on $\vec{B}$ here is a special case of the more general condition
\begin{equation}
{f^\lambda}_{[\mu} F_{\nu]\lambda} = 0
\end{equation}
found in \cite{tani}.

It remains to compute and interpret the conserved quantity $K$ that one
obtains by squaring this new supercharge. A direct computation gives the
answer  (\ref{squareK}) announced in the introduction,  where the total
angular momentum $\vec{J}$ can also be written as
\begin{eqnarray}
      \vec{J} & = & \vec{r}  \times ( \vec{p} - e \vec{A}) + \vec{S} - eg {1
\over r} \vec{r} \nonumber \\
           & = & \vec{L} - e \vec{r}  \times \vec{A}  + \vec{S} - eg {1 \over
r} \vec{r}  \, .
\end{eqnarray}
\nit
The first term $\vec{L}$ in the last formula, shows that the part of  $K$
quadratic in the momenta: ${1 \over 2}  p_i K_{ij} p_j $, is
determined by a well-known Killing-tensor
\begin{equation}
    K_{ij} = \delta_{ij} \vec{r}^{\,2} - x_i x_j   \, ,
\end{equation}
\nit
with a well-understood interpretation \cite{RH}. In fact, we write
\begin{equation}
    K_{ij}= f_{ik} \, f_{jk}
\label{added}
\end{equation}
\nit
in order to suggest an interpretation of it as a metric to which we will
return below.

We wish also to comment on our extra supersymmetry in relation to a distinct
approach to such matters, to be found in \cite{CoPa}. This requires to search
for these transformations in a superfield form. Our result can be readily cast
into such a form. Using (\ref{ourQ}) to compute $\tilde \delta\Phi_i = \left[
\tilde \epsilon \tilde Q , \Phi_i \right]$, we are soon led to the result
\begin{equation}
      \tilde \delta\Phi_i = - \epsilon \varepsilon_{ijk} D\Phi_j \Phi_k
\end{equation}
\nit
which is of the form used in \cite{CoPa} for the antisymmetric quantity
$I_{ij} =  - \varepsilon_{ijk} D\Phi_k$.
Of course, here we do not find that $I_{ij}$
defines a complex structure since we demand only invariance of the action
under $\tilde \delta$ and not closure of the supersymmetry algebra generated
by $\tilde \delta$ in the same way as in \cite{CoPa}.  Our result however
satisfies the relevant subset of conditions laid down in \cite{CoPa} for the
application in hand.
\vs

\nit
Apart from the (super)conformal symmetry that was discussed in \cite{Jackiw}
and used in \cite{dhoker}, the symmetry algebra of the magnetic monopole
therefore becomes:
\begin{equation}
\begin{array}{ccc}
   Q^2 = H, & \{ Q , \tilde Q \} = 0, &
       \tilde Q^2 = \frac{1}{2}\, \left(\vec{J}{}^2 - e^2 g^2 +
       \frac{1}{4}\right),  \\
     && \\
       \left[ Q , H \right] = 0, & [ \tilde Q , H ] = 0, & [ J_i , H ] = 0,  \\
     && \\
\left[ Q, J_i \right] = 0, & \left[ \tilde{Q}, J_i \right] = 0, &
     [J_i , J_j ] =  i \varepsilon_{ijk}  J_k.
\end{array}
\end{equation}
\nit
Such a structure in which the Killing-Yano tensor is related to a square root
of total angular momentum is familiar from other examples \cite{GRH,JW}.
It represents the particular kind of non-linearity familiar also from
finite-dimensional W-algebras \cite{BHT}.
\vs

\nit
{\bf 3.} Recently it was shown by Rietdijk and one of us  \cite{RH2}, that
there exists a certain duality which relates two theories in which the
role of the Killing-Yano tensor and the vierbein and the role of the
supercharges $Q$ and $\tilde Q$ are interchanged.
Comparison of (\ref{supe1}) and (\ref{ourQ}) indicates that similar ideas
are relevant in the theory studied here.
Thus one is led to
consider not only our original theory but also the one in which $K$ and
$\tilde Q$ are Hamiltonian and `first' supersymmetry. One knows the
Hamiltonian of the latter theory
and the canonical equations for the variables that occur in it, so
that its dynamical content can be completely determined. In fact we can
show that it describes a particle of spin $\frac{1}{2}$ confined to a sphere of
radius $\rho$, for each fixed $\rho >0$, centered at the position of a
monopole.

We consider a supersymmetric model for a particle on a sphere of
fixed radius $\rho$, in the background field of a magnetic monopole located
at the center of the sphere.  The constraints can be imposed using a
spinorial supermultiplet of Lagrange multipliers $(\lambda + \theta a)$:
\begin{equation}
\Delta L_{constr}\, =\, -\frac{a}{2}\, \left(x_i^2 - \rho^2 \right) - i \lambda
     \psi_i x_i.
\label{3.1}
\end{equation}
\nit
The full set of constraints (primary and secondary) then becomes
\begin{equation}
x_i^2 = \rho^2, \hspace{2em} x_i \psi_i = 0,  \hspace{2em} x_i \left( p_i
    - e A_i \right) = 0.
\label{3.2}
\end{equation}
\nit
After introducing Dirac-brackets, the constraints can be used to write
the classical Hamiltonian as
\hspace{2em}\begin{equation}
H^* = \frac{1}{2\rho^2}\, \left(p_i - e A_i\right)\, K_{ij}\,
\left(p_j - e A_j \right)\,-  e \, \vec{B} \cdot \vec{S}.
\label{3.3}
\end{equation}
\nit
Using (\ref{added}), it is seen that this  is equivalent to
\hspace{2em}\begin{equation}
H^* = \frac{\vec{M}^{\,2}}{2\rho^2}\, -\, e \, \vec{B} \cdot \vec{S},
\label{3.4}
\end{equation}
\nit
where $M_i=-f_{ij}\, (p_j -eA_j)=\, \varepsilon_{ijk} \, x_j (p_k-eA_k)$.
Clearly, this supersymmetric model describes a subclass of the solutions of
the original model (those with spherical symmetry), and involves precisely the
Stackel-Killing tensor as its metric and the Killing-Yano tensor as the
dreibein.

Interestingly, the modification introduced here changes the new
Killing-Yano-based supersymmetry of the original model into a standard
supersymmetry of the constrained Hamiltonian. In the quantum theory, to find
the corresponding result, we rescale the new supercharge
(\ref{ourQ}) by a factor $\rho$
\begin{equation}
\tilde{Q}\, =\, \frac{1}{\sqrt{2} \rho}\,
\left( -i \varepsilon_{ijk} \sigma_i x_j D_k + \frac{1}{2} \right),
\label{3.5}
\end{equation}
\nit
(where $D_k = \partial_k - ieA_k$), so that the square of this
supercharge for a magnetic
monopole field can be written in the form
\begin{equation}
\tilde{Q}^2\, =\frac{1}{2} \left[  -\,    D_i \left( \delta_{ij}
- \frac{x_i x_j}
{\rho^2} \right) D_j\,
   -\, e \vec{\sigma} \cdot \vec{B}\, +\, \frac{1}{4 \rho^2} \right],
\label{3.6}
\end{equation}
\nit
As a result, the supercharge (\ref{3.5}) is naturally identified with the
proper restriction of $\tilde{Q}/\rho$ of the original model to the sphere
with radius $\rho$; however, restricted to the sphere it plays the role of
the {\em ordinary} supercharge $Q$: it is a square root of the (constrained)
Hamiltonian. Thus we see, how the new non-standard supercharge of one model
is related to the standard supersymmetry of another model. Moreover, the
second model is here a constrained version of the original model.

We remark on the appearance of (\ref{3.5}) in the constrained model. In the
constrained model, the representation of the fermions $\psi_i$ and of  the
momentum $p_i$ requires an extra projection operator $K_{ij}$ in comparison
with the representation in the unconstrained model. However, the extra terms do
not contribute to the representation of $\tilde Q$ in the constrained model,
basically owing to the zero mode of the Killing-Yano tensor $f_{ij}$.

\vs\vs
\nit\hbox{{\bf Acknowledgements}\hfill}
\vs

\nit
The research of FDJ and JWvH is supported by the Human Capital and Mobility
Program through the network on {\em Constrained Dynamical Systems}\/. The
research of
KP is supported by Stichting FOM.
AJM thanks JWvH for the hospitality extended to him at NIKHEF at the time when
much of the present research was performed.


\begin{thebibliography}{99}
\bibitem{dhoker} E.\ d'Hoker and L.\ Vinet, Phys.\ Lett.\ {\bf 137B} (1984) 72.
\bibitem{Saki} M.\ Sakimoto, Phys.\ Lett.\ {\bf B 151} (1985) 115.
\bibitem{rachel} R.H.\ Rietdijk, Ph.D.\ thesis Amsterdam 1992,
        {\it Applications of supersymmetric quantum mechanics.}
\bibitem{AJM} A.J.\ Macfarlane, Nucl. Phys.\ {\bf B 438} (1995) 455.
\bibitem{martin} J.L. Martin, Proc.\ Roy.\ Soc.\  {\bf 251} (1959) 536
                 and ibid. 543.
\bibitem{GRH} G.W. Gibbons, R.H. Rietdijk and J.W. van Holten, Nucl.\ Phys.\
              {\bf B 404} (1993) 42.
\bibitem{CM} B.\ Carter and R.G.\ McLenaghan, Phys.\ Rev.\ D19 (1979), 1093.
\bibitem{KY} K.\ Yano, Ann.\ Math.\ 55 (1952), 328.
\bibitem{tani} M. Tanimoto,  TIT/HEP-277/COSMO-50, gr-qc/9501006.
\bibitem{RH}  J.W. van Holten and R.H. Rietdijk, Class.\ Quantum Grav.\
              {\bf 11 } (1993) 559.
\bibitem{CoPa} R.A.\ Coles and G.\ Papadopoulos, Class.\ Quantum Grav.\ {\bf 7}
              (1990) 427.
\bibitem{Jackiw} R.\ Jackiw, Ann.\ Phys.\ {\bf 129} (1980) 183.
\bibitem{JW} J.W.\ van Holten, Phys.\ Lett.\ {\bf B 342} (1995) 47.
\bibitem{BHT} J.\ de Boer, F.\ Harmsze and T.\ Tjin, preprint ITP-SB-95-07
              and ITFA-03-2/95.
\bibitem{RH2} R.\ Rietdijk and J.W.\ van Holten, Durham/NIKHEF preprint
               (july 1995).
\end{thebibliography}
\end{document}